\documentclass[10pt,twocolumn,twoside,a4]{article}

\usepackage{geometry}
\usepackage{graphicx}
\usepackage{url}

\begin{document}
\title{Statistical Exploration of Relationships Between Routine and Agnostic Features Towards Interpretable Risk Characterization
}

\author{Eric Wolsztynski% <-this % stops a space
\thanks{This work was presented at the 2019 IEEE Medical Imaging Conference, 2 November, Manchester, UK. This work was supported in part by Science Foundation Ireland under Grants No. 12/RC/2289-P2 and SFI-PI 11/1047.}
\thanks{E. Wolsztynski is with the School of Mathematical Sciences, University College Cork, T12 XY86, and the Insight Centre for Data Analytics, Ireland (e-mail: eric.w@ucc.ie).}}

\maketitle

\pagenumbering{gobble}

\begin{abstract}
As is typical in other fields of application of high throughput systems, radiology is faced with the challenge of interpreting increasingly sophisticated predictive models such as those derived from radiomics analyses. Interpretation may be guided by the learning output from machine learning models, which may however vary greatly with each technique. Whatever this output model, it will raise some essential questions. How do we interpret the prognostic model for clinical implementation? How can we identify potential information structures within sets of radiomic features, in order to create clinically interpretable models? And how can we recombine or exploit potential relationships between features towards improved interpretability? A number of statistical techniques are explored to assess (possibly nonlinear) relationships between radiological features from different angles.
\end{abstract}

\section{Introduction}
Building and interpretation of radiomics-based predictive models is discussed in many reports \cite{Aerts14,Soussan14,Buvat15,Desseroit16,Gillies16,Hatt17a,Hatt17b}, which all highlight the difficulty of converting the model-based risk assessment into practical decision-making pathways for routine implementation--a necessary condition to the clinical implementation of machine learning and artificial intelligence solutions in radiology. Several types of methodologies are considered in the literature to better understand potential for feature recombination towards this goal. 

Conventional models such as the lasso, random forests or neural networks \cite{Tibshirani97,Gevrey03,ESL2009,Hatt19} are usually used to build predictive models. Combined with preliminary feature elimination, these techniques provide feature set reduction methodologies geared towards a particular endpoint of interest, whether for prognosis (e.g. overall or two-year patient survival) or tumor characterization (e.g. tumor grading, subtyping, etc.) \cite{Gillies16,Asselin12,Rahim14,Alic14,Ypsilantis15,Lian16,Wolsztynski18,Wolsztynski19}. Interpretation of the model is directly derived from its structure (which describes the interaction between its covariates), and is determined in the context of the endpoint of interest. For example, a linear model for 2-year survival may indicate that an increase of 1 standard unit in both SUVmax and GLCM entropy  may contribute to a 3-fold increase in the risk of death within the next two years. In this illustration, the linear interaction between SUVmax and GLCM entropy is directly associated with worse prognosis at the 2-year horizon, but this finding provides only limited insight in terms of the role of GLCM entropy in a prognostic context or in terms of tumor characteristics. 

Microarray data analysis encompasses another family of techniques for the discovery of features with high predictive potential. In this framework, multiple statistical testing of association with endpoint is performed to select features of interest \cite{Aerts14,Gillies16,Parmar15}. Interpretation of the selected group of predictors is thus also directly linked to the endpoint of interest and may include assessment of joint association between a number of features in this context.

Nonparametric (i.e. model-free) multidimensional methods such as Principal Components Analysis (PCA) and clustering are also employed to identify relevant sets of prognostic features. These techniques are common to other high-throughput fields including genomics and proteomics \cite{Aerts14,Parmar15,Mathe16}, and consist in detecting relationships between potential predictors before their grouped association with a particular endpoint is established (this is done as a second step). They therefore provide ways to identify associations between features that are not endpoint-dependent. Their scope is however limited by their construction; for example PCA may be used to identify linear associations but not nonlinear ones \cite{Falissard99}. 

Ultimately, an output set of predictive features is considered for use in (future) clinical settings \cite{Zhao19}. For further interpretation, association of a small number of radiological features with phenotype or other clinical assessment, through e.g. logistic regression \cite{Soussan14}, can be performed. Texture and other radiomic features are sometimes clustered on the basis of correlation heat maps \cite{Gillies16}. The clinical relevance of cluster consensus maps can be assessed \cite{Parmar15}, and used to measure predictive ability of radiomic features for specific clinical, biological and functional pathways \cite{Grossmann17}. Patients may also be clustered on the basis of texture features heat maps, and availability of gene-analysis data allows for association of radiomic signature features and gene expression using gene-set enrichment analysis, by scoring radiomic signatures \cite{Aerts14}. In many studies, composite radiomic variables are defined for each patient via a linear combination of selected  features and used as additional variables alongside routine clinical or other variables \cite{Zhao19}; these can also be interpreted based e.g. on their mathematical construction.

A large number of statistical techniques are thus at hand to build predictive models and find relevant associations within feature sets. Biological interpretation of the output multivariate associations of radiologic features however remains challenging, due to the complex and diverse nature of most of these variables, and of cancer itself. Opacity of machine learning frameworks, often used as black boxes, also adds to this difficulty. They can however be used to gather insight and simplify radiomic summaries in view to facilitate further interpretation. Finding direct, statistically strong associations among features, as illustrated hereafter, can provide a mechanism to simplify such models and facilitate explainability.

\section{Methods}

\subsection{The dataset}

We consider radiological summaries derived for a set of FDG-PET sarcoma studies in a previous analysis, as reported in \cite{Wolsztynski18}.  
%%%%
This dataset of primary sarcoma tumors was acquired at the University of Washington in Seattle, United States, between August 1993 and January 2003, after patients were diagnosed by biopsy. The final cohort comprised of 197 studies, including 88 deaths before loss to follow up. The tumors consisted of 130 soft tissue, 51 bone, and 16 cartilage sarcomas, in patients aged between 17 and 86 years of age (median 45), of which 86 females and 111 males, with 99 high-grade, 66 intermediate, and 32 low-grade tumors. In this report we present the results of analyses carried out on the cohort of 130 soft tissue sarcomas (STS) from this dataset; all other subtypes have been excluded from analysis.

Quantitations were obtained for a fixed-threshold segmentation, with a threshold value set for each study based on the subsample of the lower 15\% of uptake values (so as to include background and healthy tissue activity only). For a given study, the segmentation threshold was thus defined as the mean subsample value plus three standard deviations of this subsample. Given the near-homogeneous voxel dimensions of the output images (voxel size of 4.30 mm $\times$ 4.30 mm in the transverse plane and slice thicknesses of 4.25 mm), no interpolation was performed prior to VoI resegmentation for texture analysis. Uptake values were requantized into 32 grey levels by fixed bin number transformation. 
 
A total of 43 variables were considered and may be identified in three frames: (i) routine clinical variables (tumor grade, clinical volume, patient age, patient sex, maximum standardized uptake value (SUVmax), mean uptake value (SUVmean) and total lesion glycolysis (TLG) were collected for this cohort); (ii) structural features including heterogeneity $\mathcal{H}_0$ and $\mathcal{H}_1$ as defined in \cite{Wolsztynski18}, and associated spatial uptake gradients; and (iii) a set of image summaries including morphologic and texture features, all computed as per definitions provided by the Image Biomarker Standardization Initiative (Version 1.5) \cite{IBSI,IBSI5}. More specifically, this third set of features included volume asphericity, morphological descriptors for ellipsoidal characteristics, intensity- and histogram-based first-order statistics, GLCM features, as well as two GLSZM features evaluating the numbers and sizes of contiguous homogeneous regions of equal (discretized) grey level. 

\subsection{Overall scope}

The objective here is not to propose predictive models of patient risk or tumour characterization, but rather to look for and identify patterns among features typically used in radiomic analyses. To this end a number of multivariate data exploration and modelling techniques are used as follows: 

\begin{enumerate}
\item	Correlation and partial correlation analysis, to inspect correlation structures present in the image analysis data;
\item	Multivariate decomposition and clustering, to identify natural groupings of features;
\item Regularized multilinear modeling of features, to identify small (2 or 3) subsets of features that can ``explain'' (i.e. predict) a given feature of interest. 
\end{enumerate}

We can use any of the above exploratory analyses to recombine features into composite predictive variables (based e.g. on partitional clustering techniques as in other works cited earlier), which allows for increased statistical power and data-based evaluation of model interpretability. An illustration of this step is also provided in the next section, demonstrating statistical prognostic potential of composite variables derived from these analyses on the sarcoma cohort and discussing their interpretation. 

\subsection{Correlation and partial correlation analyses}

\begin{figure}[t!]
\centering
\includegraphics[width=.7\columnwidth]{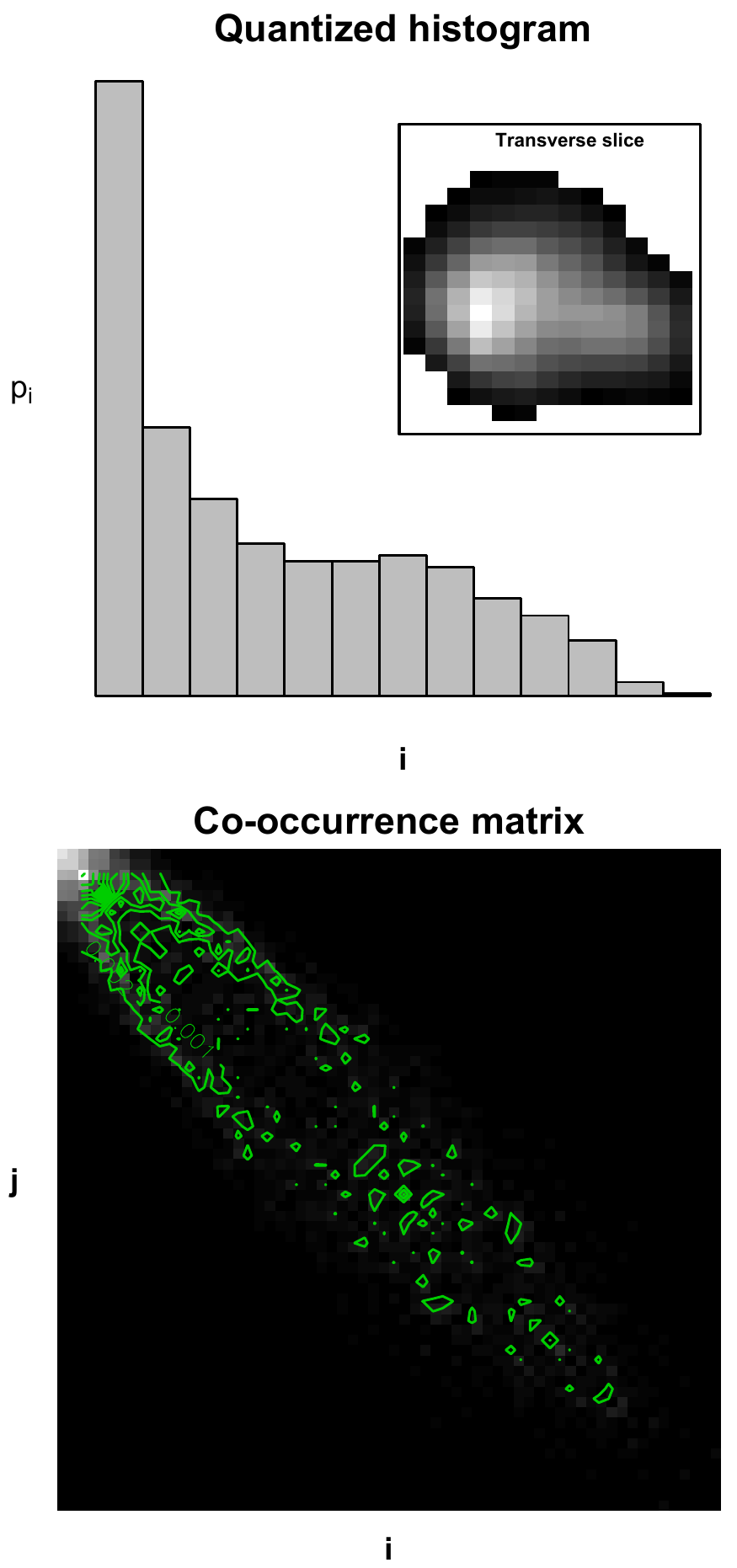}
\caption{Illustration of grey level distributions from a requantized FDG-PET sarcoma image. Inset, top: mid-volume transverse slice of the requantized FDG-PET data. Top: the histogram exhibits a clear relationship between voxel grey level $i$ and likelihood. Bottom: level curve representation of the GLCM matrix depicts the ellipsoidal footprint that is typical of joint uptake distribution structures. }
\label{fig:distributions}
\end{figure}

\begin{figure}[h!]
\centering
\includegraphics[width=.8\columnwidth]{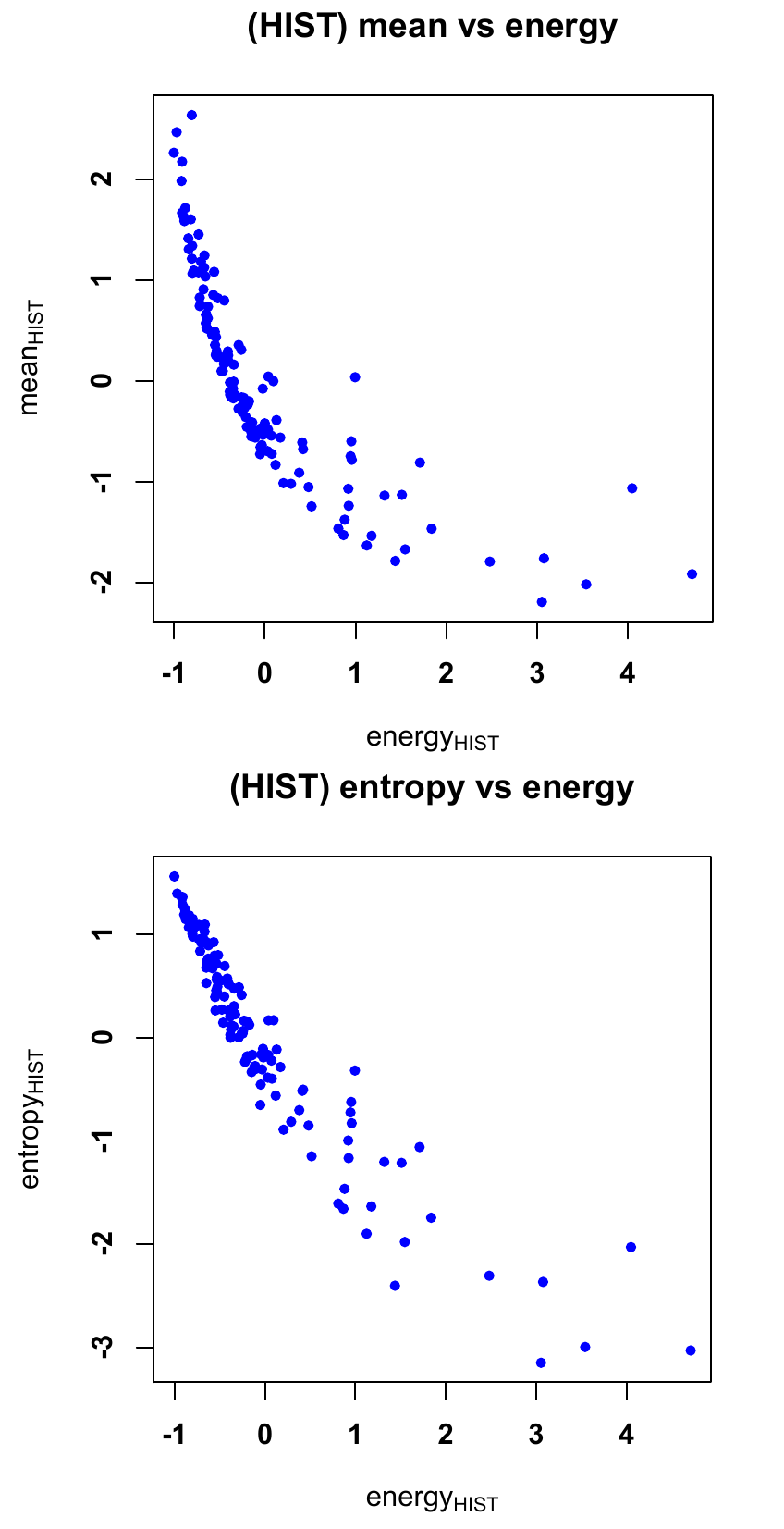}
\caption{Examples of (nonlinear) associations between texture features in the sarcoma feature set, which result from their mathematical construction.}
\label{fig:associations}
\end{figure}

Correlation may be induced by one of several causes. It may result from the mathematical construction of the features; for example it would be reasonable to think that 
$$
\mbox{mean}_{\mbox{\tiny{HIST}}} = \sum_{i=1}^{N_g} i p_i  
$$
and
$$
\mbox{energy}_{\mbox{\tiny{HIST}}} = \sum_{i=1}^{N_g} p_i^2 
$$
are closely related since  uptake histograms tend to be right-skewed, with $p_i$ decreasing as $i$ increases. The same principle applies to second-order quantitations; for example the GLCM matrix yielding joint probabilities $p_{ij}$ for voxel grey levels $i, j=1,\dots,Q$ typically exhibits a bell-shaped structure with monotonic variations in $p_{ij}$ across the probability surface. Figure \ref{fig:distributions} illustrates such structural variations in the first- and second-order distributions. Another example, considering the rough approximation $log(p_i)\approx p_i-1$ for small values of $p_i$, exposes the numerical proximity between entropy$_{HIST}$ and energy$_{HIST}$ as follows:
$$
\mbox{entropy}_{\mbox{\tiny{HIST}}} = -\sum_{i=1}^{N_g} p_i \log(p_i)  \approx -\sum_{i=1}^{N_g} p_i (p_i-1)  \approx 1 - \mbox{energy}_{\mbox{\tiny{HIST}}}
$$
Figure \ref{fig:associations} illustrates the relationships observed in the sarcoma dataset for these two examples.

In other instances, correlation may be caused e.g. by tomography-related aspects (for example relating to noise or dose levels, or the reconstruction filtering process), in which case correlation characteristics would change with scanner, or by underlying biological characteristics. Correlation analysis would not provide information on cause, but it constitutes a valuable tool in identifying associations and, therefore, potential pathways for feature set simplification.

The correlation matrix (e.g. using Pearson correlation, as done here) is often considered for exploratory purposes, but also for preliminary feature elimination. In the latter case, groups of highly correlated features are identified and a single feature is kept as unique representative for each group. How the other variables are eliminated may be guided by clinical or practical considerations but may also be, and often is, arbitrary. 

The overlap in information with other variables within the dataset, due e.g. to confounding or mathematical construction, contributes to the value of the correlation coefficient calculated between two variables. Gaussian graphical models (GGMs) \cite{Falissard99,Yuan07,Giraud09} provide a way of analysing partial correlation between variables instead. GGMs can be used to highlight direct relationships (the edges in the graph, as illustrated in Figure \ref{fig:GGM}) between features that are conditionally dependent given all other variables. Dependence is evaluated here as a non-negligible partial correlation between the two features. 

\subsection{Multivariate exploration}

Here we considered Principal Component Analysis (PCA) for multidimensional exploration of the feature set. It also consists in analyzing the correlation structure of the feature set, but provides a more elaborate tool for assessment of multivariate associations. Note that PCA applies to quantitative features (we therefore excluded categorical features from the analysis), but variations on this approach may be used for mixed quantitative and descriptive feature sets, allowing for inclusion of categorical agnostic variables. 

Partition-based clustering was applied as in \cite{Wolsztynski18} to the PCA projection matrix (i.e. the matrix of eigenvectors obtained from spectral decomposition of the feature set correlation matrix) in order to identify groupings of variables in the feature set. The features projected via PCA can be used directly as composite predictor variables, as detailed for example in \cite{Wolsztynski18}. These linear recombinations can however be difficult to interpret. Subsequent clustering analysis of feature groupings in the information space can be exploited to recreate alternative composite variables with  meaningful clinical interpretation.

\subsection{Multilinear analysis of features}

The above methodologies are mechanisms used to isolate groupings of variables based on the correlation (or partial correlation) structure of the feature set. Direct association of features can also be identified and exploited via multivariate modelling, using one feature as the dependent variable explained by subset of other features. Here we considered multilinear modelling and used lasso models to identify a group of 2 or 3 features in order to describe (i.e. predict) each feature in the dataset.

$N_{test}$=30 observations were randomly taken out of the original STS dataset for use as an independent test set. Repeated 5-fold cross-validation (CV), using two repetitions, was applied to the remaining $N_{CV}$=100 observations, which was performed using each one of the continuous variables in the STS feature successively as the dependent variable, and all other variables as predictors in a lasso model. (No preliminary feature elimination was performed.) In total P=41 lasso models of the form (for some i.i.d. zero-mean Gaussian noise $\varepsilon$)
\begin{equation}\label{eq:lasso}
X_j = \beta_0+\beta_1X^{(-j)}_1 + \dots + +\beta_{P-1}X^{(-j)}_{P-1} +\varepsilon
\end{equation}
were therefore fitted to each feature of interest $X_j$ via CV using the remaining P-1 features $\{X^{(-j)}_{1},\dots,X^{(-j)}_{P-1}\}$ as predictors, from the CV sample of $N_{CV}$ observations, i.e. $X_j=\{X_{j,1}, \dots, X_{j,N_{CV}}\}$. In other words each of the 41 continuous features available (that is, all features except for tumor grade and patient sex) was thus modelled by fitting a lasso model to the remainder of the feature set. 

Feature selections and model fits from the CV training sets were then analysed. The feature selection scheme provided by lasso was used to eliminate weaker contributions. The remaining predictors were inspected and those with an estimated effect $\hat{\beta}$ of a magnitude of at least 20\% of the overall sum of estimated effects $\sum_{j=1}^{P-1} \hat{\beta}_j$ from (\ref{eq:lasso}) were retained as final predictors of that dependent variable. The choice of a 20\% cutoff was arbitrary but aimed to reduce the model to a few (possibly strong) predictors. In any case (i.e. even when all estimated coefficients were under this cutoff point of 20\% of the cumulative effect in magnitude) the two covariates with highest estimated effect $\hat{\beta}$ were retained as a final model. 

A final lasso model was fitted to the whole CV sample (using all $N_{CV}=100$ observations) using only these (typically two or three) most popular features as determined by CV. This final prediction model was then applied independently to the test set (using the $N_{test}=30$ remaining observations) for final prediction performance assessment for each of the features available. 

\section{Results}

\subsection{Correlation and partial correlation analyses}

\begin{figure}[h!]
\centering
\includegraphics[width=\columnwidth]{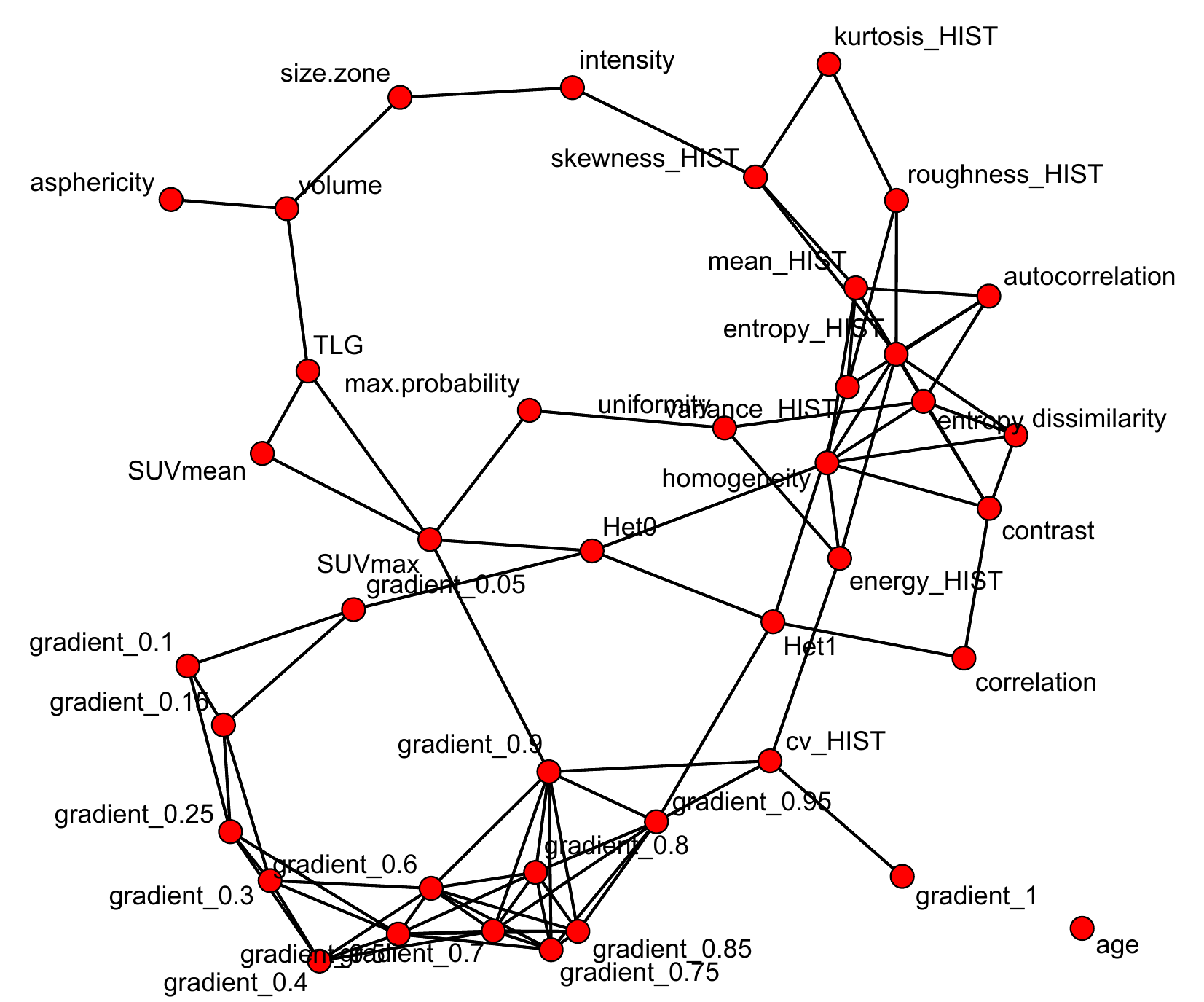}
\caption{Gaussian graphical model obtained for the sarcoma feature set, showing direct associations between features as defined by partial correlation.}
\label{fig:GGM}
\end{figure}

Figure \ref{fig:GGM} illustrates the output of a GGM, which provides guidance for the understanding of direct relationships between some of the features based on their partial correlations. It highlights several characteristics of this feature space. For instance, it indicates no direct correlation between age and any of the features and exhibits the expected relationships between SUVmax and SUVmean, and between SUVmean, TLG and volume. It also exhibits a cluster of uptake gradients (lower left) which was also expected due to their construction--these quantities are successive quantiles of the sample of normalized uptake gradients \cite{Wolsztynski18}. Moreover this graph provides information on direct associations between some of the conventional radiomic features. Skewness, kurtosis and other summaries of the histogram of requantized intensities, unsurprisingly, tend to cluster together (top right), with significant partial correlation found among this group of features. The graph also highlights direct correlation among a group of GLCM features comprising of entropy, dissimilarity, contrast, homogeneity and a few other metrics. 

These direct associations may be at least partially explained by their mathematical construction; however they may also be partially driven by other factors that may be pertaining to biological or physiological aspects of the disease. Many of these relationships can be directly exploited in further multivariate modelling of the features to assess prediction potential for this set of features, as considered further below. 

\subsection{Principle Component Analysis}

\begin{figure}[h!]
\centering
\includegraphics[width=.7\columnwidth]{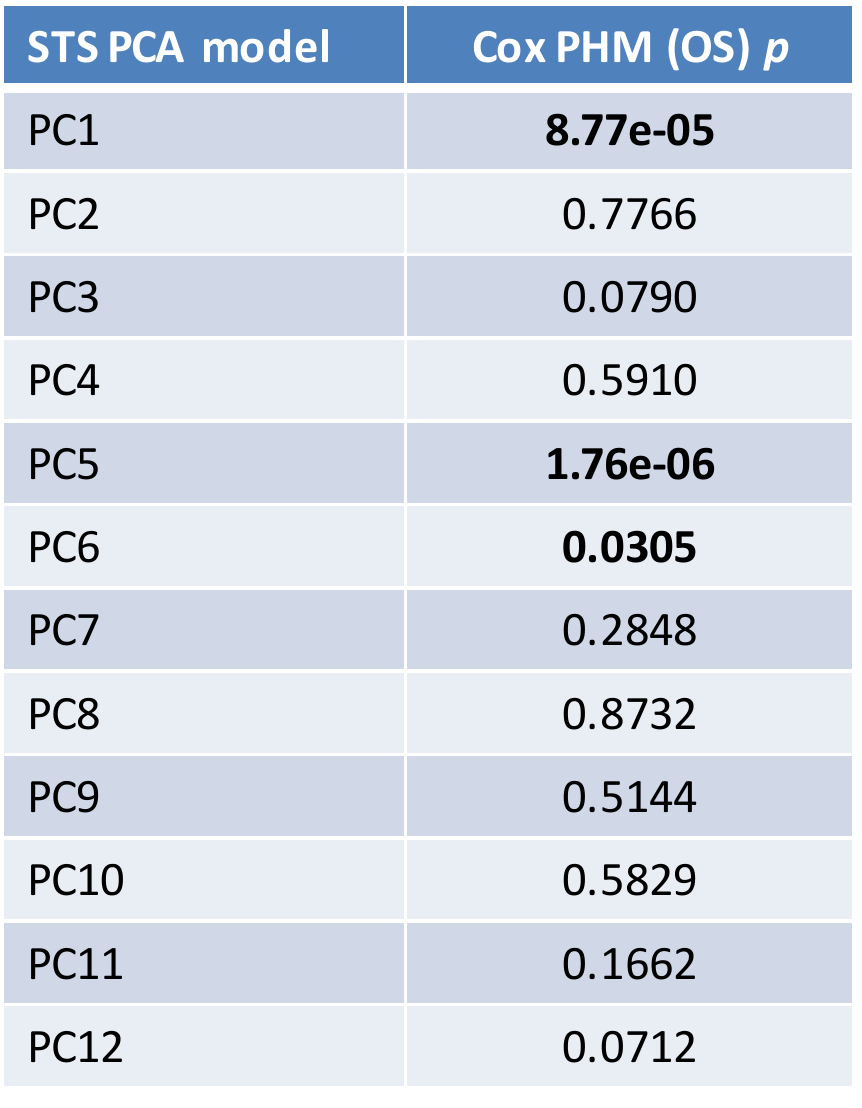}
\caption{P-values for the multivariate Cox proportional hazard model for overall survival copmrising of the frst 12 principal components, with associated concordance index $\mathcal{C}$=0.75.}
\label{tab:PCA_Cox}
\end{figure}

\begin{figure}[h!]
\centering
\includegraphics[width=\columnwidth]{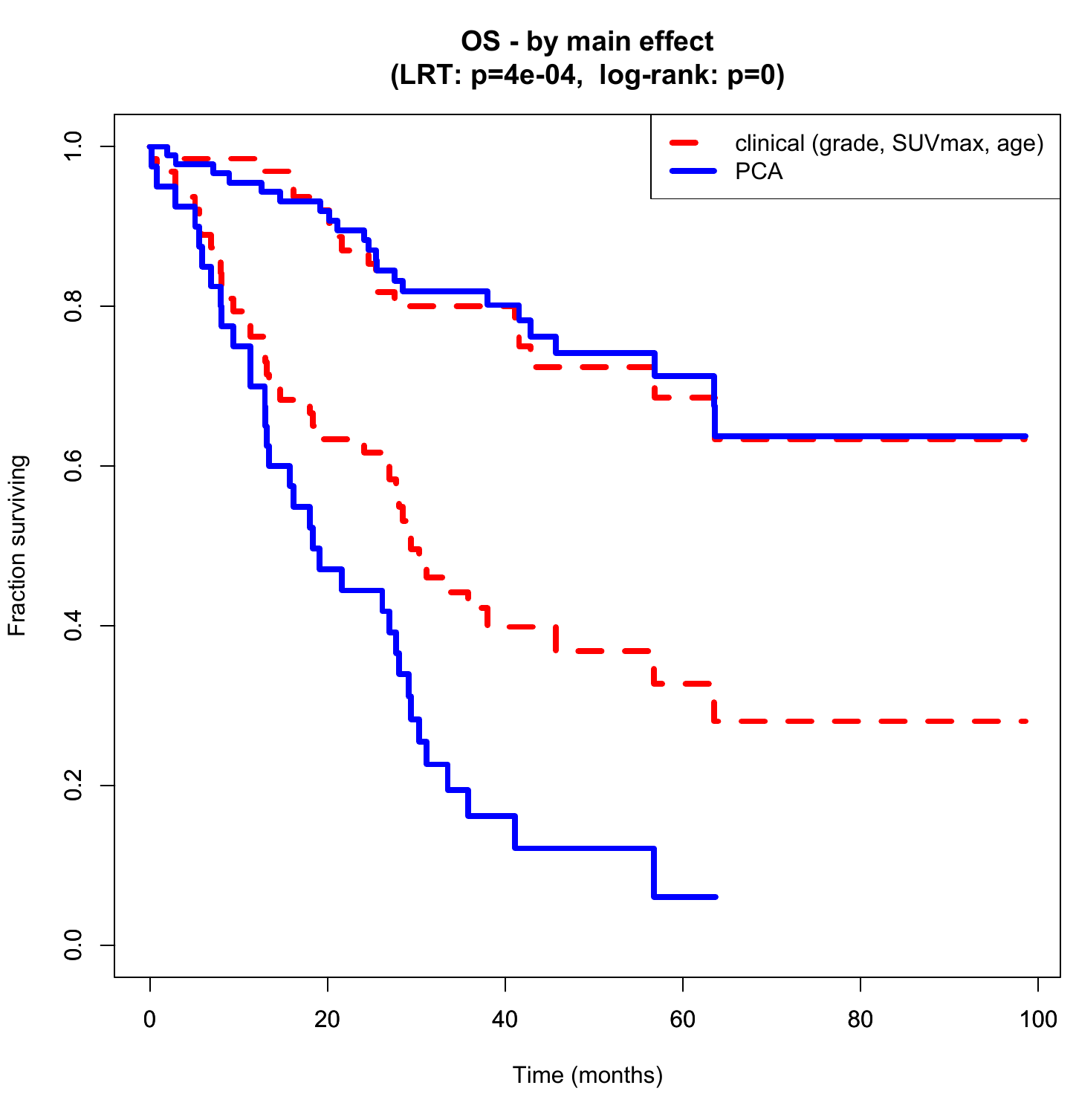}
\caption{Kaplan-Meier analysis showing risk stratifications obtained from baseline clinical assessment based on grade, age and SUVmax (red, dashed lines) and by using the first 12 principal components together for multivarate risk prediction (blue, solid lines).}
\label{fig:PCA_KM}
\end{figure}

The principle components (PC's) derived from PCA of the STS feature set consist of linear recombinations of the input features. The first 12 PC's (on the arbitrary basis that the first 12 PCs captured over 95\% of the variance in the feature set) were used as composite risk predictors in a multivariate Cox proportional hazard model. This analysis demonstrated the prognostic potential of this multilinear recombination of the original STS feature set, with 3 of the 12 PC's found statistically significant prognostic variables at the 5\% significance level (Figure \ref{tab:PCA_Cox}). Figure \ref{fig:PCA_KM} further illustrates this potential in terms of Kaplan-Meier risk stratification for overall survival, comparing this model to a baseline clinical model comprising of tumor grade, patient age and SUVmax. Note that here the low- and high-risk survival curves are separated so as to optimize the log-rank test statistic.

\begin{figure}[h!]
\centering
\includegraphics[width=\columnwidth]{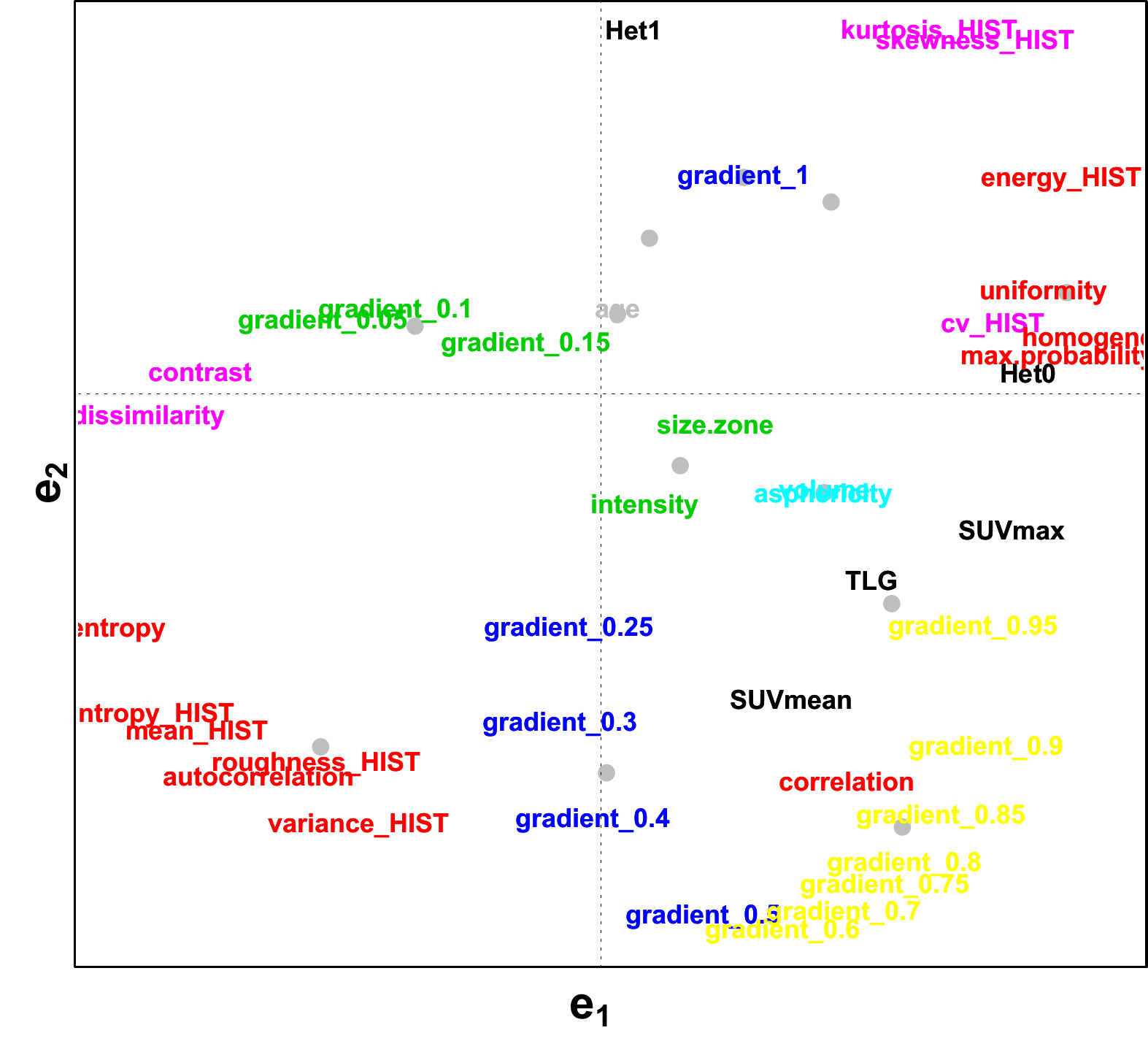}
\caption{Biplot of the features in the PCA projection space, along the first two eigenvectors ($e_1$ and $e_2$). Colour-code and grey dots respectively indicate clusters and cluster centroids obtained from k-means clustering of the projected features (i.e. analysing the eigenvector coordinates).}
\label{fig:PCA}
\end{figure}

\begin{figure}[h!]
\centering
\includegraphics[width=\columnwidth]{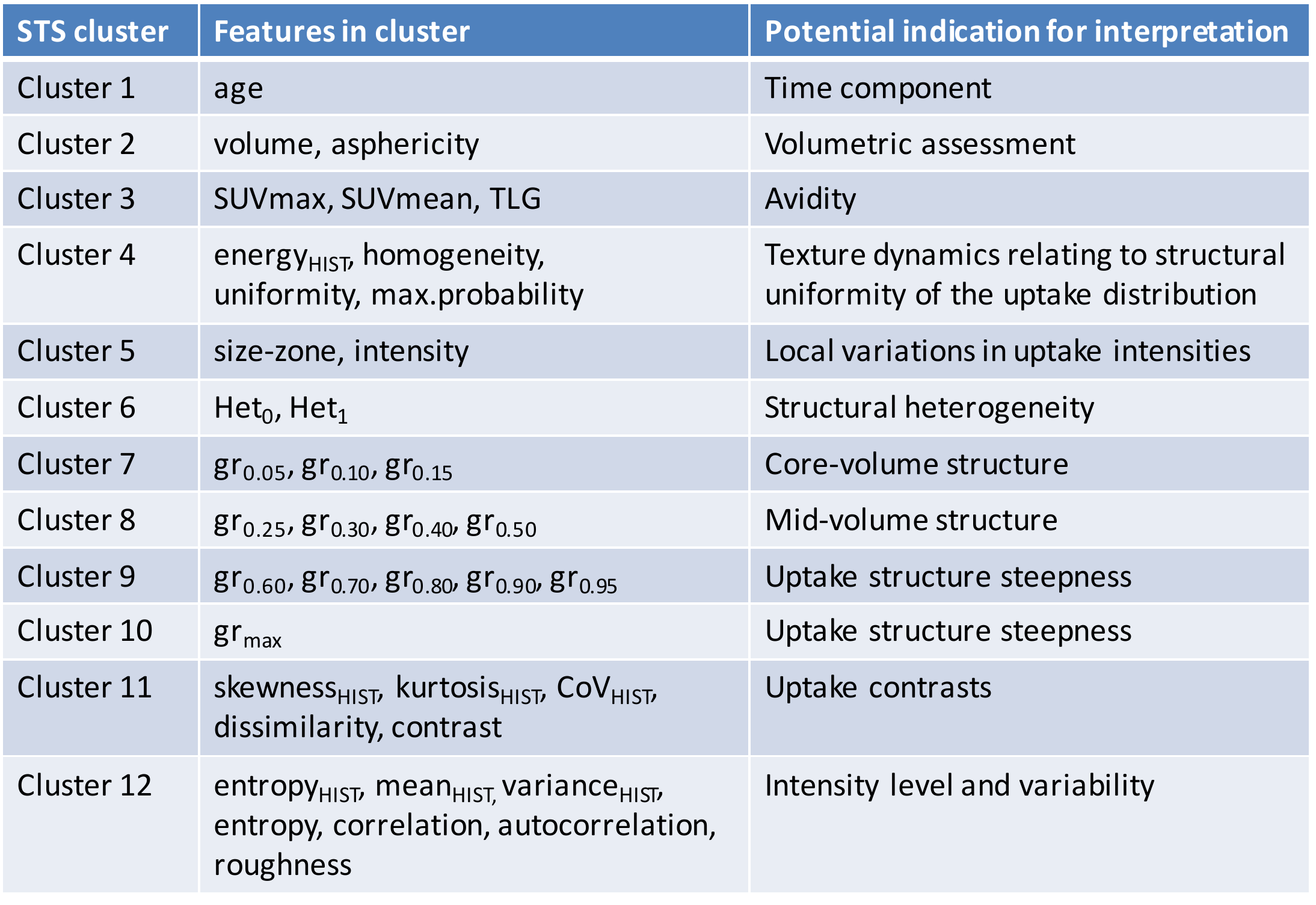}
\caption{Clusters of features obtained from combined PCA and k-means analysis of the STS feature set. A tentative indication for interpretability is provided here only for the purpose of illustrating the potential for explainable PCA-derived models facilitated by the clustering analysis.}
\label{tab:PCA_clusters}
\end{figure}

This PCA output was subsequently analysed by k-means clustering using 12 clusters (thus arbitrarily matching our use of the first 12 PCs for risk prediction). This determined the clusters of features in the PCA projection space illustrated by Figure \ref{fig:PCA}. Makeup of the clusters is detailed in Figure \ref{tab:PCA_clusters}. The right-most column in this table provides tentative fields of interpretability for each cluster. This is included here solely as an illustration of the potential for explainable PCA-derived models that is facilitated by clustering analysis of the PCA output. Thorough, rigorous investigations would be required in order to establish suitable clinical interpretation guidelines from such feature clusters.

\subsection{Multilinear modelling}

\begin{figure}[h!]
\centering
\includegraphics[width=\columnwidth]{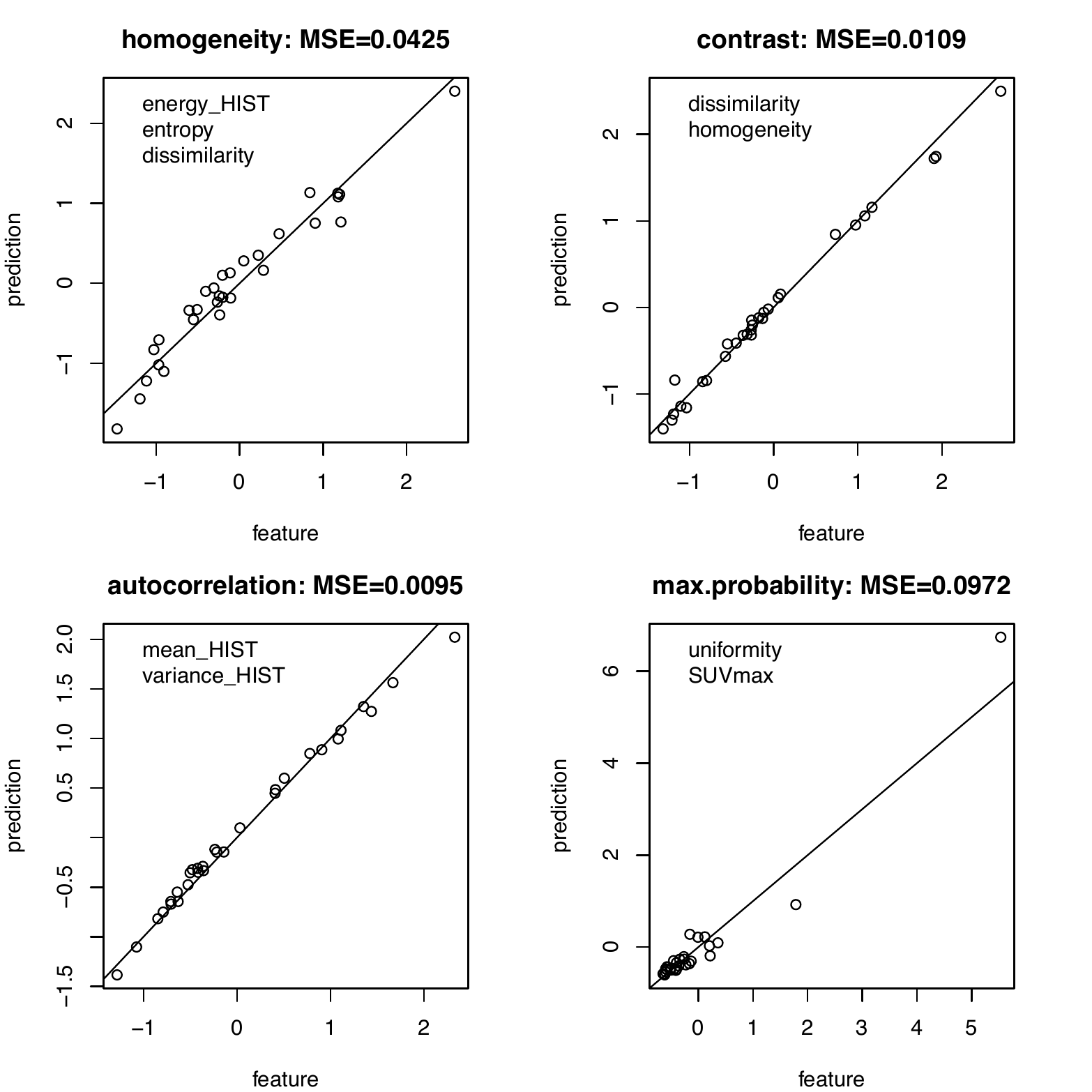}
\caption{Prediction performance based on lasso modelling for homogeneity, contrast, autocorrelation and maximum probability (all GLCM features) successively (clockwise from top-left), with associated MSE between predicted and observed values, for the STS test set ($N_{test}$=30). The two or three features used as predictors in each case are indicated in inset.}
\label{fig:associations1}
\end{figure}

\begin{figure}[h!]
\centering
\includegraphics[width=\columnwidth]{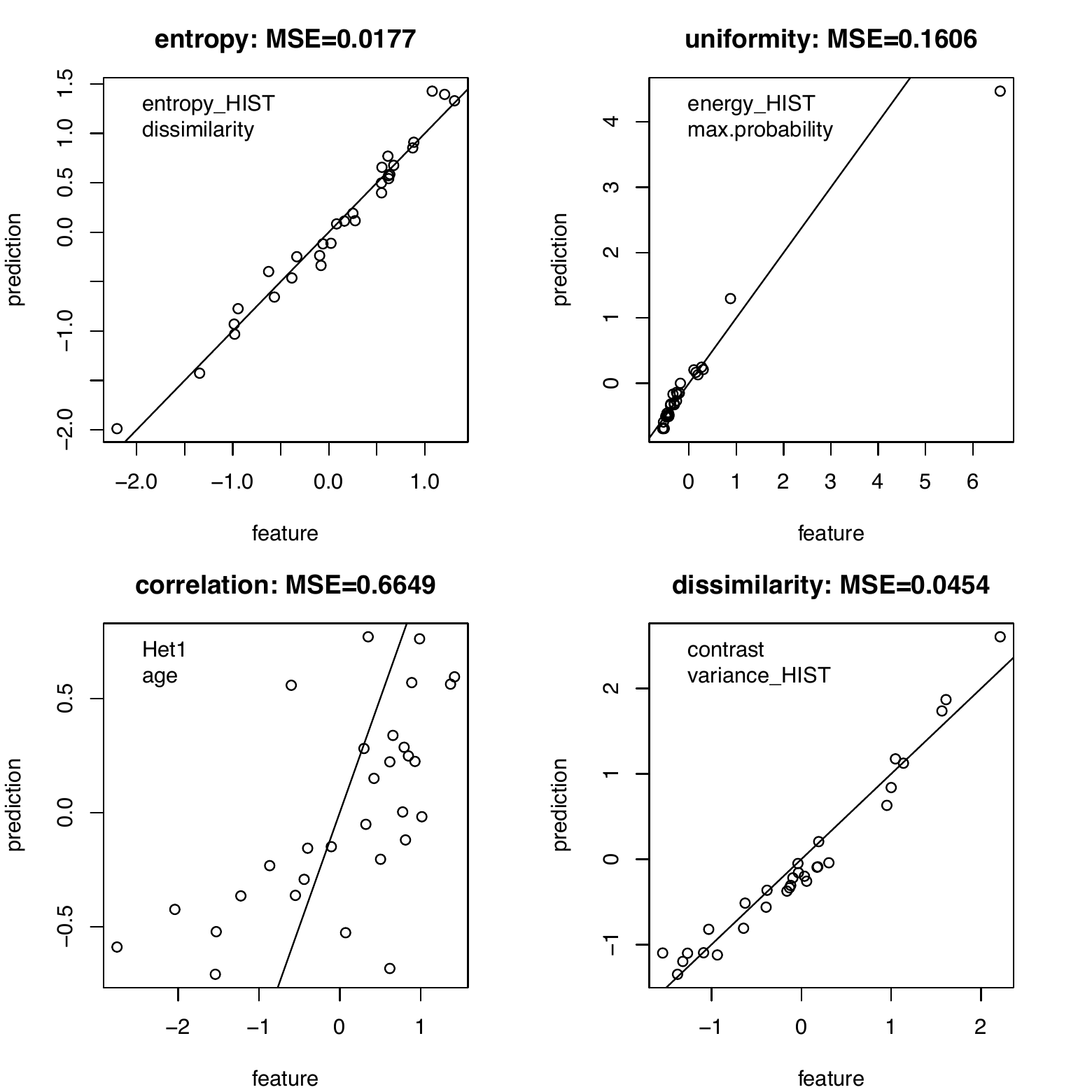}
\caption{Prediction performance as in Figure \ref{fig:associations1}, but here for entropy, uniformity correlation and dissimilarity GLCM features respectively.}
\label{fig:associations2}
\end{figure}

Lasso modelling of the features determined that in many cases, a radiomic feature could be predicted with high accuracy. Figures \ref{fig:associations1} and \ref{fig:associations2} show eight examples of such predictions on the 30 test datapoints. GLCM autocorrelation, for example, was predicted extremely well using only mean$_{HIST}$ and variance$_{HIST}$, which suggests this second-order feature may be replaced with, or interpreted by more easily explainable first-order features. 

In some cases clinical variables were found useful in predicting agnostic features; for example Figure \ref{fig:associations1} depicts reasonable prediction of GLCM max.probability using GLCM uniformity and SUVmax. 

We also note that age and $\mathcal{H}_1$ were selected to predict GLCM correlation however with poor performance (Figure \ref{fig:associations2}). This aligns with previous findings (from the above correlation analysis and also from \cite{Wolsztynski18}) of reasonable separation between model-derived features ($\mathcal{H}_0$, $\mathcal{H}_1$ and related uptake gradients) and conventional radiomic features. Overall, however, many of the 41 features considered were predicted with high accuracy from a small number of other features. 

For many of these features, it was observed that the predictors selected via cross-validated lasso modelling were connected to the dependent feature in the GGM representation; in other words lasso often selected predictors with strong partial correlation with the dependent variable (which could be expected from this multilinear modelling technique).

\section{Conclusion}

Identifying and understanding associations between radiomic features and routine variables, such as grade or other clinically interpretable forms of assessment, is a key step towards integration of models obtained from machine learning analyses. Some works do establish links between specific texture features and tumor biologic heterogeneity for example; but the generalization of this understanding to all such agnostic features will enable integration of full AI systems within the field of radiology. Here we promote discussion on some possible statistical analysis pathways towards this goal. 

Most reports of radiomics studies focus on the clustering of agnostic features and assess their potential  alignment with routine variables \textit{a posteriori}, to analyze the clustering output. Potential interactions between agnostic and routine variable can also be explored more directly by allowing any such feature to describe any other, as is done in this work via lasso modelling.

This pilot analysis highlighted numerous substantial feature associations. Many of 41 clinical and radiomic features considered here were predicted with high accuracy using a small number of other features, via lasso modelling, focusing on linear interactions. The majority of these associations were noticed to coincide with groupings of features obtained on the basis of a high partial correlation, as exposed in a Gaussian graphical model representation. 

This didactic presentation aimed at demonstrating methodologies of interest for exploration of strong direct associations within a radiological feature set. More extensive analyses of such associations are currently underway and will be presented in follow-on reports for sarcoma, NSCLC and other cancer types. This work will explore opportunities for simplification of prognostic models on the basis of relationships found within the combined clinical and agnostic feature set.

\bibliographystyle{IEEEtran}
\bibliography{bib_pet.bib}

\end{document}